\documentclass[twocolumn,showpacs,preprintnumbers,amsmath,amssymb]{revtex4}
\usepackage{graphics}
\usepackage{epsfig}
\begin{document}
\title{Quantum phase transitions in rotating nuclei}
\author{R.~G.~ Nazmitdinov}
\affiliation{Departament de F{\'\i}sica,
Universitat de les Illes Balears, E-07122 Palma de Mallorca, Spain}
\affiliation{Bogoliubov Laboratory of Theoretical Physics,
Joint Institute for Nuclear Research, 141980 Dubna, Russia}
\author{J. Kvasil}
\affiliation{Institute of Particle and Nuclear Physics, Charles
University, V.Hole\v sovi\v ck\'ach 2, CZ-18000 Praha 8, Czech Republic}

\begin{abstract}
We extend the classical Landau theory for rotating nuclei and  show
that the backbending in $^{162}Yb$,  that comes about as a result
of the two-quasiparticle alignment, is identified with the second order phase
transition. We found  that the backbending in $^{156}Dy$, caused by the
instability of $\gamma$-vibrations in the rotating frame, corresponds to
the first order phase transition.
\end{abstract}
\maketitle



Quantum phase transitions (QPTs), that occur at zero temperature as a function of
some nonthermal control parameter, are studied extensively in diverse areas of
physics including, for example, low-dimensional systems in
condensed matter physics \cite{Sad}, atomic nuclei and molecules \cite{Jo02}.
In nuclear structure, QPTs associated with transitions between different shapes
or geometric configurations of a chain of nuclei in the ground state are
being studied intensively during the past years
within the interacting boson model (cf \cite{Jo02}).
Thanks to novel experimental detectors, a new frontier of discrete-line
$\gamma$-spectroscopy at very high spins has been opened in the rare-earth nuclei.
These nuclei can accommodate the highest values of the angular
momentum, providing one with various nuclear structure phenomena.
Evidently, from analysis of the rotational states
one expects to obtain further insight into
the nature of QPTs in finite quantum systems.

\begin{figure*}[ht]
\includegraphics[height=.3\textheight]{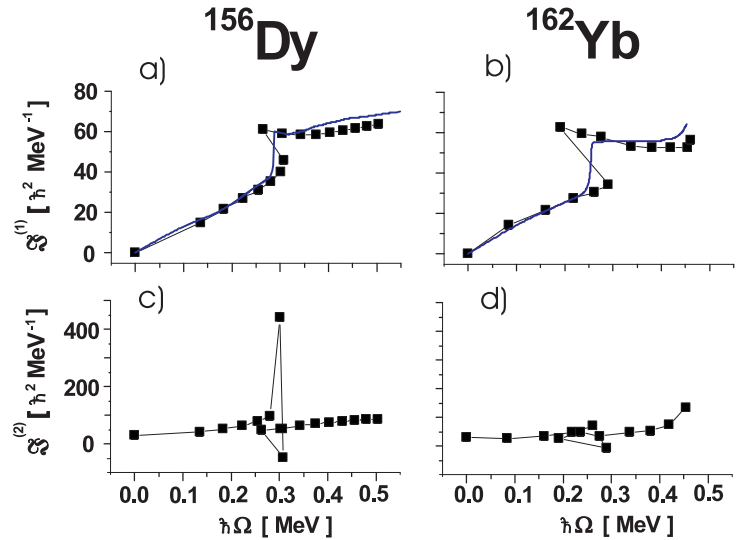}
\caption{
Rotational behavior of the experimental, kinematical
$\Im^{(1)}=I/\Omega$ and
dynamical $\Im^{(2)}\approx 4/\Delta E_{\gamma}$ moments of inertia.
Here, $\hbar\Omega=E_{\gamma}/2$, $E_{\gamma}$
is the $\gamma$-transition energy between two neighboring states that differ on
two units of the angular momentum $I$ and $\Delta E_{\gamma}$ is the difference
between two consecutive $\gamma$-transitions. The experimental data denoted by
black squares are taken from \cite{brook}. The experimental rotational
frequency at the transition point is $\hbar\Omega_c \approx 0.27, 0.32$ MeV for
$^{162}Yb$ and $^{156}Dy$, respectively.  The results of calculations
for $\Im^{(1)}$ are connected by a solid line.}
\label{fig1}
\end{figure*}

 One of the well known phenomena at high spins  is a backbending.
A sudden increase of
a nuclear kinematical moment of inertia ${\Im}^{(1)}=I/\Omega$ of the lowest (yrast) level
sequence as a function of a rotational frequency $\Omega$
(see experimental results for $^{156}Dy$ and $^{162}Yb$ in Fig.\ref{fig1})
is a paradigm of structural changes in a nucleus under rotation.
While one observes a similar picture for the backbending
in the considered nuclei (see Fig.\ref{fig1}a,b), a different response of a nuclear field
upon the rotation becomes more evident with the aid of the experimental dynamical moment
of inertia $\Im^{(2)}=dI/d\Omega\approx4/\Delta E_\gamma$ as a function of the
angular frequency (see Fig.\ref{fig1} c, d).
Indeed, the dynamical moment of inertia, due to the
relation $\Im^{(2)} = \Im^{(1)}+\Omega d\Im^{(1)}/d\Omega$,
is very sensitive to structural changes of a nuclear field.
At the transition point $\Im^{(2)}$ wildly fluctuates with a huge amplitude
in $^{156}Dy$,  whereas these fluctuation are quite mild in $^{162}Yb$.
One of the purposes of the present contribution is  to interpret the experimental
results in $^{156}Dy$ and $^{162}Yb$ (see Fig.\ref{fig1})
from perspective of the Landau theory of phase transitions within a
microscopic approach \cite{JR} based on  the cranked
Nilsson model plus random phase approximation (CRPA).
We will show that the interplay between the alignment (single-particle degrees of freedom)
and collective quantum  fluctuations  determines the type of
the shape-phase transition at the backbending.

The phase transition is usually detected by means of an order parameter as
a function of a control parameter.  In rotating nuclei one can suggest
a few order parameters like deformation parameters of a nuclear
effective potential, $\beta$ and $\gamma$, that characterize the geometrical
configuration (cf \cite{Jo02}), - as a function of the rotational frequency,
i.e., the control parameter.
To analyze the experimental data of above nuclei, we use a
cranked Hamiltonian
\begin{equation}
H_{\Omega} = H - \sum_{\tau =N,P} \lambda_{\tau}\hat N_{\tau}-\,
\Omega \hat J_x+\, H_{\rm int}.
\label{h1}
\end{equation}
The term  $H=H_{\it Nil}\,+\, H_{\rm add}$ contains the Nilsson Hamiltonian
$H_{\it Nil}$ and the additional term that restores the local Galilean invariance
of the Nilsson potential in the rotating frame.
The Nilsson potential naturally incorporates the quadrupole deformation parameters
($\beta$ and $\gamma$) of a nuclear shape. The interaction includes
separable monopole pairing, double stretched
quadrupole-quadrupole (QQ) and monopole-monopole terms. The details about
the model Hamiltonian (\ref{h1}) can be found in \cite{JR,wob}.
In our approach mean field parameters are determined from the
energy-minimization procedure (see \cite{JR}).
The consistency between the mean field and the residual interactions
of the Hamiltonian (\ref{h1}) was achieved by varying the strength
constants of the pairing and $QQ$ interactions in the CRPA.
It results in the separation collective excitations
from those that are related to the symmetries broken by the mean field.
Among them are the conservation of particle numbers and space
symmetries (see details in \cite{JR,wob}).

We found that the triaxiality of the mean field sets
in at the rotational frequency  $\hbar \Omega_{c}=0.250$, $0.301$ MeV
for $^{162}Yb$, $^{156}Dy$, respectively,
which triggers a backbending in the considered nuclei (see Fig.1 in \cite{JR}).
To elucidate the different character of the shape
transition from axially symmetric to the triaxial shape and its relation to
a phase transition, we consider
potential landscape sections in the vicinity of the shape transition.
Since we analyze a shape transition from the axially symmetric shape
($\gamma = 0$) to the triaxial one ($\gamma \neq 0$), we choose only
the deformation parameter $\gamma$ as the order parameter
that reflects the broken axial symmetry.
Such a choice is well justified, since the deformation parameter $\beta$
preserves its value before and after the shape transition in both nuclei:
$\beta_t \approx 0.2$ for $^{162}Yb$ and
$\beta_t \approx 0.31$ for $^{156}Dy$. Thus, we consider
a mean field value of the cranking Hamiltonian,
$E_{\Omega}(\gamma;\beta_t) \equiv \langle H_{\Omega} \rangle$, for different
values of $\Omega$ (our state variable) and $\gamma$ (order parameter)
at fixed value of $\beta_t$.

\begin{figure*}[ht]
\includegraphics[height=.3\textheight]{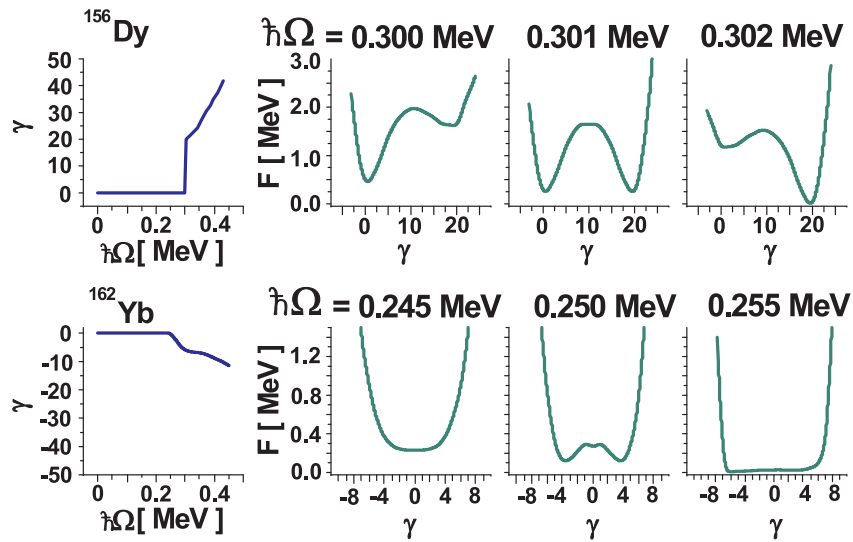}
\caption{
The rotational dependence of the order parameter $\gamma$ and
the energy surfaces sections  $F(\Omega,\gamma)=E_{\Omega}(\gamma;\beta_t)-E_{min}$
for $^{156}Dy$ (top) and $^{162}Yb$ (bottom) before and after the transition point.
The energy is given relative  to the value $E_{min}=E_{\Omega}(\beta_t,\gamma)$ at
 $\hbar\Omega=0.255,\,\,0.302$ MeV for $^{162}Yb$, $^{156}Dy$,
respectively.
}
\label{phatr}
\end{figure*}

For $^{156}Dy$ we observe the emergence of the order parameter $\gamma$
above the critical value $\hbar\Omega_c=0.301$ MeV
of the control parameter $\Omega$ (see a top panel in Fig.\ref{phatr}).
Below and above the transition point there is a unique phase  whose properties
are continuously connected to one of the coexistent phases at the transition point.
The order parameter changes discontinuously as the nucleus passes through the
critical point from axially symmetric shape to the triaxial one.
The polynomial fit of the potential landscape section
at $\hbar\Omega_c=0.301$ MeV yields the following expression
\begin{equation}
F(\Omega;\gamma) =F_0(\Omega)+ F_2(\Omega) \gamma^2 - F_3(\Omega) \gamma^3 +
F_4(\Omega) \gamma^4 ,
\end{equation}
where the coefficients $F_0(\Omega)=0.3169$ MeV, $\gamma$ in degrees and
$F_2(\Omega)=0.12239$, $F_3(\Omega)=0.009199$,
$F_4(\Omega)=1.7\times10^{-4}$ are defined in  corresponding units.
We can transform this polynomial to the form
\begin{equation}
{\bar F}=\frac{F(\Omega;\gamma) -F_0(\Omega)}{\bar {F_0}}
\approx \alpha \frac{\eta^2}{2}- \frac{\eta^3}{3} +\frac{\eta^4}{4}
\label{gen}
\end{equation}
where ${\bar {F_0}}=(3F_3)^4/(4F_4)^3$,
$\alpha=8F_2F_4/(9F_3^2)$, $\eta=4F_4/(3F_3)\gamma$.
The expression (\ref{gen}) represents the generic form of  the anharmonic model
of {\it the structural first order phase
transitions} in condensed matter physics.
The condition $\partial {\bar F}/\partial \eta=0$ determines the following
solutions for the order parameter $\eta$ :
$\eta_1=0$, $\eta_{2,3}=(1\pm\sqrt{1-4\alpha})/2$.
If $\alpha>1/4$, the functional ${\bar F}$ has a single minimum at $\eta=0$.
Depending on  values of $\alpha$, defined in the interval
$0<\alpha<1/4$, the functional ${\bar F}$ manifests the transition
from one stable minimum at zero order parameter via one minimum+metastable state
to the other stable minimum with the nonzero order parameter.
In particular, at the universal value of $\alpha=2/9$ the functional
${\bar F}$ has two minimum values with
${\bar F}=0$ at $\eta=0\rightarrow \gamma\approx0^0$ and
$\eta=2/3\rightarrow \gamma \approx27^0$
and a maximum at $\eta=1/3\rightarrow \gamma\approx13.5^0$. The correspondence
between the actual value $\gamma\approx 20^0$ and the one obtained from the
generic model  is quite good.

In general, the alignment of angular
momenta of a nucleon pair occupying a high-j intruder orbital near the Fermi surface
is considered as a main driving force that leads to the backbending.
Although two-quasiparticle
states align their angular momenta along the axis x (collective rotation),
the axial symmetry persists till the transition point
(see also discussion in \cite{JR}). The RPA analysis of $\gamma$-vibrational
(lowest) excitations of the positive
signature in the vicinity of the shape transition
demonstrates a collective nature of these excitations.The mode blocks a transition to
the triaxial shape. However, at the transition point,
this mode is anomalously low in the rotating frame. It appears that
soft positive signature $\gamma$-vibrations (fluctuations of
the order parameter $\gamma$) coupled to the other modes are
responsible for the shape-phase transition of the first order.
A drastic change of the mean field configuration leads to large fluctuations of
the dynamical moment of inertia at the transition point, since
$\Im^{(2)}=-d^2E_{\Omega}/d^2\Omega$, which is  reproduced successfully in
the CRPA (see details in \cite{JR}).
It  seems reasonable to say that the backbending in $^{156}Dy$ possesses typical
features of {\it the first order phase transition}.

In the case of $^{162}Yb$ the energy $E_{\Omega}(\gamma;\beta_t)$  and
the order parameter (Fig.\ref{phatr}) are smooth  functions in the vicinity of
the transition point $\Omega_c$. This implies that two phases,
$\gamma=0$ and $\gamma\neq0$, on either side of the transition point should
coincide.
Therefore, for $\Omega$ near the transition point $\Omega_c$
we can expand our functional
$F(\Omega,\gamma)=E_{\Omega}(\gamma,\beta_t)-E_{min}$
in the form
\begin{equation}
F(\Omega;\gamma) = F_1(\Omega) \gamma +
F_2(\Omega) \gamma^2 + F_3(\Omega) \gamma^3 + F_4(\Omega) \gamma^4 + \ldots
\label{5}
\end{equation}
The energy surfaces are symmetric with regard of
the sign of $\gamma$ and this also supports the idea that the effective energy $F$
can be expressed as an analytic function of the order parameter $\gamma$.
With the aid of the conditions of the phase equilibrium,
${\partial F}/{\partial \gamma}=0$ and ${\partial^2 F}/{\partial \gamma^2}\geq 0$,
one can show that $F_1(\Omega_c)=F_2(\Omega_c)=0$. In virtue of these equations
and of the fact that
all phases at the transition point should coincide, we obtain from
${\partial F}/{\partial \gamma}=0$  that
$F_3(\Omega = \Omega_c) = 0$.
Assuming that $F_3=0$ for all $\Omega$, the minimum condition
 ${\partial F}/{\partial \gamma}=0$ yields
the following solution for the order parameter
\begin{equation}
\gamma_1=0 \,, \quad
\gamma_{2,3}^2 = - \frac{F_2(\Omega)}{2\,F_4(\Omega)} =
\left\{
\begin{array}{ll}
\neq 0 & for \,\, \Omega\neq \Omega_c \\
= 0    & for \,\, \Omega = \Omega_c
\end{array}
\right.
\label{18}
\end{equation}
Since at the transition point $F_2(\Omega_c)=0$, one can propose
the following definition of the function $F_2(\Omega)$:
\begin{equation}
F_2(\Omega) \approx \frac{dF(\Omega;\gamma)}{d\Omega}\, \left(\Omega - \Omega_c \right)
\label{19}
\end{equation}
Thus,  we have $\gamma\sim (\Omega-\Omega_c)^\nu$ and the critical exponent
$\nu=1/2$, in accord with the classical Landau theory, where the temperature is
replaced by the rotational frequency.
Indeed, the numerical results (see Fig.\ref{phatr}) are
in an agreement with Eqs.(\ref{18}),(\ref{19}):
$\gamma =0$ for $\hbar \Omega < \hbar \Omega_c $, while
$\gamma \neq 0$ for $\hbar \Omega > \hbar \Omega_c $.
Thus, the backbending in  $^{162}Yb$ can be classified as {\it the phase transition
of the second order}. The smooth behavior of the function $F$ at the transition
point implies a
small amplitude of fluctuations of the dynamical moment of inertia.
This result is nicely reproduced within the CRPA approach \cite{JR}.
The RPA analysis of the lowest $\gamma$-vibrational mode in
$^{162}Yb$ indicates on the breakdown of the quadrupole phonon.
At the vicinity of the transition point
one proton and one neutron two-quasiparticle components dominate ($\sim 95\%$)
in the phonon structure and the backbending is caused by
the alignment of the neutron two-quasiparticle configuration.

The identification of wobbling (negative signature) excitations near the yrast line
provides   sure evidence of the onset of the triaxiality.
We found (see \cite{wob}) that a
shape-phase transition produces relatively high-lying wobbling
vibrational states in $^{156}Dy$. In contrast, a soft
shape-phase transition from the axially
deformed to nonaxial shapes  produces the low-lying wobbling excitations in $^{162}Yb$.
We have also established the relation between the sign of the $\gamma$-deformation and
selection rules for the quadrupole electric transitions from the wobbling to the yrast
states (see \cite{wob}). From our calculations it follows that at low angular momenta
($\hbar \Omega \leq 0.28,0.3$ MeV in $^{162}Yb$, $^{156}Dy$, respectively)
the first negative signature one-phonon band populates with approximately equal
probabilities the yrast states with $I^\prime=I\pm1$ ($I$ is the angular momentum
of the excited state). At $\hbar \Omega_c$ a shape-phase transition
occurs, that leads to the triaxial shapes with the negative $\gamma$-deformation
in the both nuclei.
In turn, the negative signature phonon band decays stronger on the yrast states with
angular momenta $I^\prime=I-1$, starting from $\hbar\Omega\geq\hbar \Omega_c$.
We predict also the dominance of $\Delta I=1 \hbar$
magnetic transitions from the wobbling to the yrast states, independently
from the sign of the $\gamma$-deformation (see \cite{wob}).


\section*{Acknowledgements}
This work was partly supported
by Votruba-Blokhintcev program of BLTP (JINR),
Grant No. FIS2005-02796 (MEC, Spain) and
Grant of RFBR No. 08-02-00118 (Russia).



\begin{thebibliography}{9}

\bibitem{Sad}
                S.~Sachdev,
                \emph{Quantum Phase Transitions},
                Cambridge University Press, Cambridge, 1999.

\bibitem{Jo02}
                J.~Jolie, {\it et al.,}
                \emph{Phys. Rev. Letters} \textbf{89}, 182502 (2002);
                F.~Iachello,
                \emph{ibid} \textbf{91}, 132502 (2003);
                F.~Iachello and N.~V.  Zamfir, {\it ibid}
                \textbf{92}, 212501 (2004).

\bibitem{brook}
                http://www.nndc.bnl.gov/nudat2/
\bibitem{JR}
                J.~Kvasil and R.~G. Nazmitdinov,
                \emph{Phys. Rev. C} \textbf{73}, 014312 (2006).

\bibitem{wob}
                R.~G. Nazmitdinov and J.~Kvasil,
                \emph{JETP} \textbf{105}, 962 (2007);
                \emph{Phys. Lett. B} \textbf{\bf 650}, 331 (2007).
\end{thebibliography}
\end{document}